# Simulation de traces réelles d'E/S disque de PC

Jalil Boukhobza, Claude Timsit

Université de Versailles Saint Quentin en Yvelines
PRiSM, Bat Descartes, 45 Avenue des Etats-Unis
78000 Versailles - France
{Jalil.boukhobza, claude.timsit}@prism.uvsq.fr

**Résumé**

Sous Windows, les outils de benchmarking des Entrées/Sorties (E/S) existants ne suffisent pas pour permettre à un développeur de définir efficacement sa stratégie d'accès aux fichiers d'après un ensemble de contraintes d'application. Ceci est essentiellement dû au fait que ces outils ne permettent de tester qu'un ensemble réduit de charges d'E/S qui ne correspondent généralement pas à ceux de l'application ciblée. Pour palier ce problème, nous avons conçu et développé un simulateur d'E/S très précis permettant de simuler n'importe quelle trace d'E/S réelle sur une architecture donnée, et dans lequel, la plupart des stratégies de cache du système de fichiers et du disque, leurs interactions, ainsi que le comportement détaillé du disque, sont simulés. Les résultats des simulations sur différentes charges d'E/S et architectures montrent un très haut degré de précision. En effet, un taux d'erreurs moyen de 6% par rapport aux mesures a été observé avec un maximum local de 10% sur les débits globaux. Nous présentons dans cet article un simulateur d'E/S disque sous Windows permettant une évaluation très précise des performances de traces réelles d'E/S. Il permet d'identifier la meilleure stratégie d'accès aux fichiers pour une trace d'E/S donnée. Ce simulateur est très facile à paramétrer grâce à l'outil d'extraction de paramètres que nous avons précédemment développé.

**Mots-clés** : simulateur d'E/S, performances des E/S, benchmarking, cache du système de fichiers, cache du disque, disque.

## 1. Introduction

Les applications multimédias, et plus généralement les applications à E/S intensives sont de plus en plus présentes et sont souvent limitées par les performances du système de stockage [12][13][14]. L'utilisation d'une stratégie d'accès aux fichiers adéquate et spécifique à l'application et à l'architecture choisie est cruciale pour assurer un fonctionnement optimal de ce type d'application.
Bien que Windows soit l'un des systèmes d'exploitation les plus utilisés, le comportement de son système d'E/S est peu étudié. S'il est possible de trouver dans la littérature une description architecturale de ce système [3] [4] [5] en plus de certains concepts clés utilisés, aucune indication n'est donnée quant aux performances produites et aux politiques de cache utilisées. C'est pour cette raison que nous nous sommes intéressés à l'analyse et la compréhension des performances d'E/S de ce type de système [30].
Malgré le fait que le système de stockage soit l'un des composants les plus lents d'un PC, ce dernier est très souvent très mal exploité. En effet, le système d'exploitation Windows procure plusieurs options quant à l'utilisation du cache du système de fichiers. Ces options correspondent à l'activation de certains drapeaux dans la fonction d'ouverture et de création de fichiers *CreateFile*. Un autre paramètre provoquant des variations de performance très importantes est la taille des requêtes de lecture/écriture. Beaucoup de développeurs ne prennent malheureusement pas en compte la totalité de ces paramètres. Ils se contentent souvent d'utiliser le mode par défaut, allant jusqu'à utiliser des tailles de requête aléatoires dans certains cas. Et lorsqu'une chute de performance est observée, le système de stockage est souvent blâmé alors qu'avec une bonne stratégie d'accès aux fichiers, la performance aurait été meilleure. Les causes exactes de ces chutes de performances sont très souvent méconnues.

Par exemple, lors d'accès séquentiels sur une machine donnée, le seul fait de changer de mode d'accès peut provoquer une importante chute de performance extrêmement pénalisante d'un facteur 10 (4Mo/s pour un disque pouvant soutenir 40Mo/sec) ou plus fréquemment des chutes d'un facteur 2 voire 3 [30]. De manière analogue, le seul fait de changer la taille de la requête de lecture/écriture peut entraîner des chutes de performances d'un facteur 2. Dire que « plus la taille des requêtes est importante et plus les performances d'E/S sont bonnes » n'est plus toujours vrai.

Pour éviter ces chutes de performances, nous avons conçu et développé un simulateur d'E/S sous Windows. Il permet de prédire très finement les performances d'une charge d'E/S dans le but d'identifier la meilleure stratégie d'accès aux fichiers pour l'application cible. Pour un architecte, le simulateur permet de mieux dimensionner le système de stockage d'après des charges d'E/S données. Cet outil simule les comportements spécifiques à des versions récentes du système d'exploitation Windows (2000, XP et 2003).

Un travail d'ingénierie inversée conséquent a été effectué pour permettre l'identification des différentes politiques de cache utilisées par le système d'exploitation et par les différents contrôleurs des disques testés. Ce sont les résultats de ces études qui ont permis la construction du simulateur d'E/S que nous décrivons dans cet article.

La section 2 de l'article décrit les travaux précédemment réalisés. La section 3 donne un aperçu du cheminement des requêtes d'accès aux fichiers sous le système d'exploitation Windows. La section 4 détaille le simulateur que nous avons développé. La dernière section contient un résumé de l'article et les perspectives envisagées.

**2. Travaux précédemment réalisés**

Quelques travaux concernant des mesures de performance globale sur des systèmes Windows ont été précédemment réalisés [1] [2]. Les travaux de Riedel et al. [1] consistent en des mesures de performance des systèmes de stockage sous WindowsNT4 sur des technologies et des architectures de stockage différentes. Les métriques de performance collectées sont les débits globaux, les latences et taux d'utilisation du processeur ainsi que l'activité du bus d'E/S. Cette étude a l'avantage d'analyser un vaste spectre d'architectures différentes mais se limite à la présentation des variations des performances *globales* d'accès aux fichiers, c'est-à-dire les débits. Les travaux de Chung et al. [2] constituent une suite des travaux précédents, sur le système Windows2000 et sur des machines plus récentes. Des comparaisons ont été effectuées entre Windows2000 et WindowsNT4. Plusieurs technologies ont fait l'objet d'expérimentations et de mesures dans cette étude. Pour les systèmes mono disque, différentes interfaces ont été testées ainsi que certaines architectures RAID dans le cas multi disques.

Cependant, les études présentées dans [1] et [2] n'expliquent pas les chutes de performances observées et ne vont pas jusqu'à étudier les temps de réponse des requêtes pour analyser plus finement les comportements des systèmes de stockage testés. Si ces études sont exhaustives pour les architectures testées, elles restent relativement superficielles pour les analyses effectuées.

De notre côté, nous avons analysé finement le système de stockage sur quelques architectures [30] [31]. Nos études sur les temps de réponse nous ont démontré un même comportement du système d'exploitation relatif aux modes d'accès utilisés quelque soit la machine testée (cf. section3). Par comportement, nous voulons dire : les algorithmes de cache du système de fichiers, l'ordonnancement ainsi que la concaténation et la division des requêtes de lecture/écriture de l'application. Ces comportements interagissent fortement avec ceux des caches des disques de la machine testée produisant des temps de réponse *périodiques* spécifiques à la machine testée et à la charge de travail appliquée.

Sous Unix, plusieurs travaux ont été réalisés dans ce domaine : parmi les plus importants, un simulateur de système de stockage nommé Disksim [7] et un outil d'extraction de paramètres pour les disques SCSI [11] permettant de configurer ce simulateur. Cet outil simule l'ensemble des comportements mécaniques du disque en plus de différents algorithmes de cache [10] [16] [20] et permet une définition flexible des générateurs de requêtes. D'autres simulateurs existent pour les systèmes Unix tel que Pantheon [21], [19] et [18]. Il n'existe aucun outil de ce type, pour les systèmes de stockage Windows, qui implémente les comportements spécifiques à ce système, en particulier les algorithmes de gestion du cache du système de fichiers. Windows et Unix présentent des architectures de système de stockage très différentes. C'est pour cette raison que nous avons construit un nouveau simulateur permettant de prédire les performances des accès aux fichiers sous Windows et prenant en compte toutes les spécificités du système Windows.

**3. Aperçu sur le cheminement des requêtes d'accès aux fichiers sous Windows**

Cette section résume très brièvement le fonctionnement des accès aux fichiers sous Windows [3] [4].

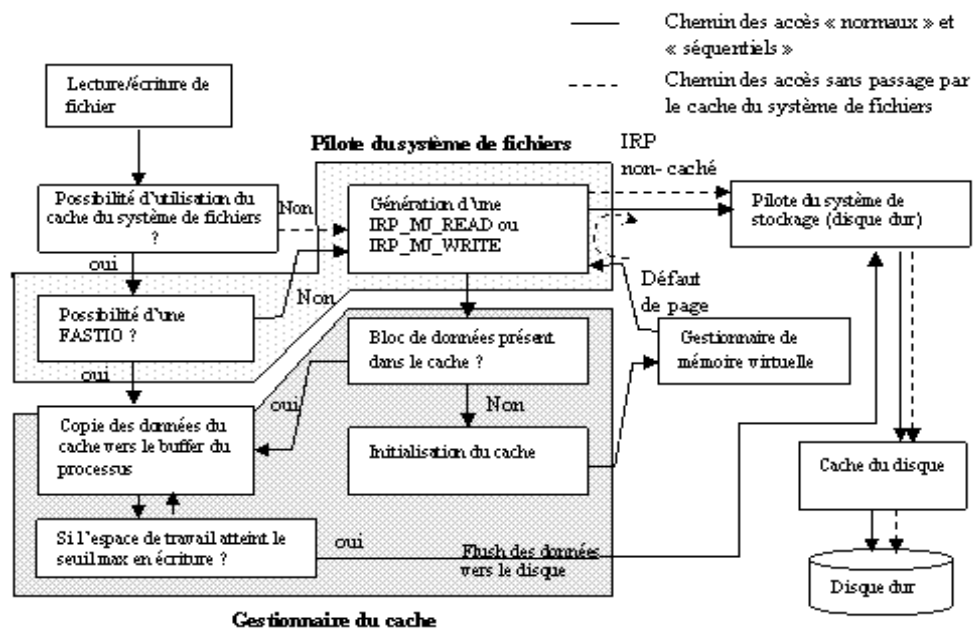

**FIG. 1** – Fonctionnement des lectures / écritures sous Windows2000

Lorsqu'un développeur veut accéder aux données d'un fichier, il passe par les fonctions d'E/S Win32 *CreateFile()*, *ReadFile()*, *WriteFile()* et *CloseHandle()*. La fonction *CreateFile()* est utilisée pour l'ouverture et la création de fichiers. On peut spécifier, grâce à certains drapeaux la manière avec laquelle on veut accéder au fichier pour aider le gestionnaire du cache à mieux optimiser les accès en choisissant (ou supprimant), par exemple, l'algorithme de préchargement (pour les lectures) ou celui d'écriture retardée (pour les écritures). Si l'on doit accéder séquentiellement aux blocs d'un fichier, il existe un drapeau (FILE_FLAG_SEQUENTIAL_SCAN) que l'on peut activer pour indiquer au système la méthode d'accès suivie pour qu'il puisse optimiser les performances.
Quatre modes d'accès ont ainsi été testés :
   o   un mode où le passage par le cache du système de fichiers est désactivé : mode « sans buffer » et « write through » pour les écritures,
   o   et deux modes pour lesquels ce dernier est activé : mode « normal » et « séquentiel ».
Les données peuvent, bien évidemment, être stockées sur le disque, dans le cache du disque ou dans le cache du système de fichiers (« cache logiciel » en mémoire principale).

**3.1. Désactivation du passage par le cache du système de fichiers**
La désactivation du passage par le cache du système de fichiers est réalisée par l'ajout dans la fonction *CreateFile()* du drapeau FILE_FLAG_NO_BUFFERING. Une opération d'appel à la fonction *ReadFile()* pour la lecture d'un fichier invoque la fonction *NTReadFile()*. Si le *handle* passé en paramètre dans la fonction *ReadFile()* indique que l'appelant a obtenu la permission de lire le fichier à l'ouverture, *NtReadFile* crée un IRP (IO Request Packet) du type IRP_MJ_READ (lecture), et l'envoie au Pilote du système de fichiers où le fichier à lire est stocké (cf. FIG. 1). Ce dernier envoie une requête directement vers le pilote du périphérique de stockage (dans notre cas un disque) en court-circuitant le gestionnaire de cache du système de fichiers et finalement charge les données du disque vers l'espace mémoire du processus via DMA. Pour l'écriture, le chemin est similaire, si ce n'est le nom de la fonction appelée – *NtWriteFile*-, et celui de l'IRP généré, IRP_MJ_WRITE.

**3.2. Activation du passage par le cache du système de fichiers**
Les blocs demandés par les fonctions utilisant les deux modes de lecture « normal » et « séquentiel » passent par le même chemin de données (cf. FIG. 1). Les fonctions *NTReadFile()* ou *NTWriteFile()* invoquées envoient d'abord une requête *FASTIO* avant de générer un IRP. Une *FASTIO* est une requête « plus légère » envoyée directement au gestionnaire de cache. Dans le cas où la donnée y serait disponible, elle est directement copiée du cache vers l'espace mémoire du processus qui la demande.
Si la requête *FASTIO* renvoie un échec, un IRP_MJ_READ (ou IRP_MJ_WRITE pour l'écriture) est crée et envoyé au pilote du système de fichiers où se trouve le fichier demandé. Ce dernier vérifie que le mécanisme de cache du système de fichiers est activé pour ce fichier. Si les données ne sont pas disponibles sur la mémoire physique, un défaut de page est engendré et envoyé vers le gestionnaire de mémoire. Ce dernier se charge de remédier au défaut de page par la création d'un IRP qu'il envoie au pilote du système de fichiers mais, cette fois-ci, l'IRP est marqué comme étant « non-caché » (absent du cache) pour indiquer au système qu'il doit lire la donnée directement sur le disque. Cette donnée est alors mise dans le cache du système de fichiers et par la suite copiée dans la mémoire du processus appelant.

**4. Le simulateur d'E/S développé**

Les outils de benchmarking d'E/S ne suffisent pas pour un développeur à optimiser une stratégie d'accès aux fichiers. Néanmoins, certains de ces outils donnent une idée globale relativement précise des performances du système de stockage. L'idéal serait, bien sûr, un outil alliant la flexibilité des simulations et la précision des résultats des outils de mesure. Nous avons essayé de nous rapprocher de ce modèle en effectuant un travail exhaustif d'ingénierie inversée sur les résultats des mesures des différents outils de benchmarking pour comprendre le comportement du système de stockage [31], puis nous avons implémenté les comportements inférés dans notre simulateur.

**4.1 L'architecture du simulateur**
L'architecture globale du simulateur est représentée dans la FIG. 2. Elle est constituée d'un ensemble de modules indépendants communicant via des mécanismes de passage de message. Le simulateur à été construit sous Omnet++ [6], un environnement de simulation libre, très flexible et pourvu d'une très bonne interface graphique. Il permet de concevoir des simulateurs basés sur des techniques de passage de messages. Cette section décrit brièvement les différents modules.

**4.1.1 Les générateurs de requêtes d'E/S**
Les générateurs de requêtes ou les fichiers de traces (cf. section 4.1.2) sont le point de départ de la simulation. L'utilisateur peut facilement définir le nombre de générateurs nécessaires pour modéliser les applications à simuler. Ceci est exprimé en termes de type d'opération (lecture, écriture, ouverture ou fermeture de fichier), temps d'inter arrivée entre les requêtes, taille des requêtes, mode d'accès et adresses logiques des blocs de données accédés. Il est possible de définir les paramètres précédemment cités par des valeurs constantes ou alors d'utiliser les différentes distributions fournies : uniforme, exponentielle, normale, binomiale, Poisson ou encore aléatoire, etc. Ceci rend la simulation très souple en nous permettant de simuler un large spectre de configurations.

**4.1.2 Les fichiers de trace d'E/S**
On peut choisir deux types d'entrées dans notre simulateur : des requêtes créées par un générateur très flexible ou des traces réelles collectées avec l'outil Filemon [23]. Le fichier de trace résultant est formaté par un outil que nous avons développé lui permettant d'être compatible avec le simulateur. Le traceur récupère plusieurs informations dont la date d'envoi de la requête, le processus concerné, le type de requête, le fichier accédé, le bloc de données demandé et le statut de la requête :

```
80       17:03:26.407    testwrite.exe:928    OPEN    C:\1\testwrite.exe    SUCCESS Options: Open  Access: Execute
85       17:03:26.407    csrss.exe:712        OPEN    C:\1\testwrite.exe    SUCCESS Options: Open  Access: All
88       17:3:26.407     csrss.exe:712        READ    C:\1\testwrite.exe    LCN: 403019       Offset: 0   Length: 12
```

| | | | | | | |
|---|---|---|---|---|---|---|
| 91 | 17:03:26.407 | csrss.exe:712 | CLOSE | C:\1\testwrite.exe | SUCCESS | |
| 95 | 17:03:26.417 | explorer.exe:2044 | OPEN | C:\1\testwrite.exe | SUCCESS | Options: Open  Access: All |
| 98 | 17:3:26.417 | explorer.exe:2044 | READ | C:\1\testwrite.exe | LCN: 403019 | Offset: 0  Length: 12 |
| 104 | 17:03:26.427 | testwrite.exe:928 | OPEN | C:\1\results\result_perf.xls | SUCCESS | Options: OpenIf  Access: All |
| 132 | 17:03:26.427 | testwrite.exe:928 | OPEN | C:\1\results\result_resp.xls | SUCCESS | Options: OpenIf  Access: All |
| 159 | 17:03:26.437 | testwrite.exe:928 | OPEN | C:\1\testwrite0 | SUCCESS | Options: Open NoBuffer Access: All |
| 190 | 17:3:26.437 | testwrite.exe:928 | WRITE | C:\1\testwrite0 | LCN: 2000668 | Offset: 0  Length: 196608 |
| 191 | 17:3:26.437 | testwrite.exe:928 | WRITE | C:\1\testwrite0 | LCN: 2000692 | Offset: 0  Length: 131072 |

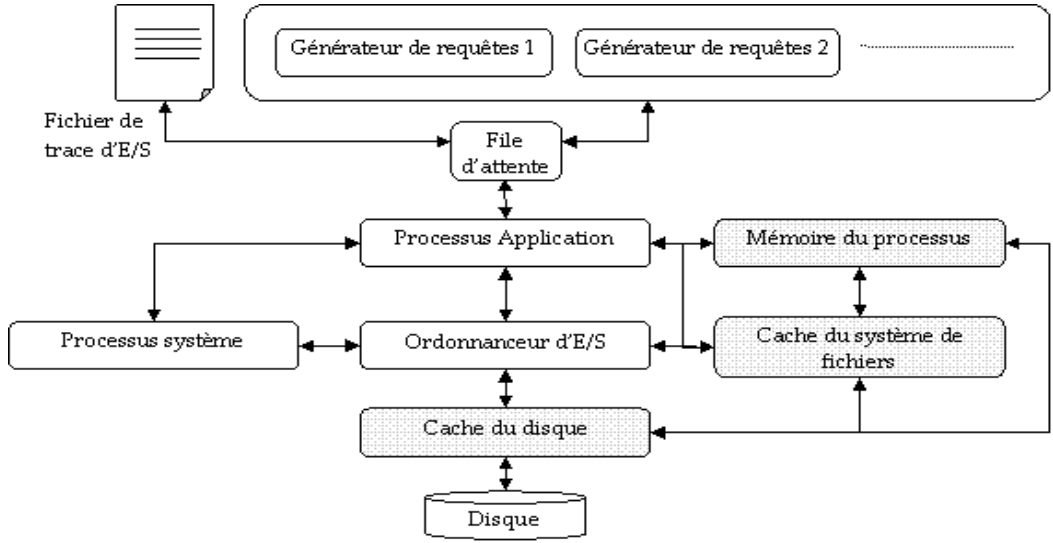

**FIG. 2 –** Architecture du simulateur d'E/S sous Windows

### 4.1.3 Le processus application et la mémoire du processus
Ces modules simulent les comportements du processus application et sa mémoire. Lorsqu'une requête est envoyée par un générateur donné, elle peut être traitée par le processus application ou le processus système [31]. Le processus système satisfait la requête, puis copie les données du cache du système de fichiers vers la mémoire du processus. Pour les requêtes dont les données ne doivent pas être stockées dans le cache du système de fichiers (drapeau FILE_FLAG_NO_BUFFERING activé), les données sont directement copiées du cache du disque vers la mémoire du processus (via DMA dans les systèmes réels) par le processus application, le processus système n'ayant aucun rôle dans ce cas.

### 4.1.4 Le processus système et le cache du système de fichiers
Ces modules simulent l'activité du cache du système de fichiers des systèmes Windows testés. Son comportement est relatif au mode d'accès et à la taille de la requête et ne dépend pas de l'architecture physique utilisée (par exemple disque SCSI ou IDE). Le nombre de blocs préchargés sur un fichier, la détection des caractéristiques de la séquence de requêtes et la taille du bloc de préchargement sont des exemples de paramètres spécifiques à un mode d'accès donné. L'une des particularités du cache du système de fichiers est qu'il n'intervient, généralement, que par bloc de 64Ko. Ceci veut dire que pour les requêtes dont la taille y est inférieure, 64Ko de données sont chargés, et si la requête a une taille supérieure à 64Ko, elle sera subdivisée en plusieurs blocs qui seront traités séparément. Nous verrons quelques détails des algorithmes utilisés plus loin.

Un autre point très important est que même si le temps d'inter arrivée entre les requêtes est nul, le système interagit avec ces dernières et par conséquent, il modifie la séquence de requêtes perçu par le sous système de stockage par rapport à celle envoyée par l'application.

Ces deux modules sont évidemment inactifs lors de l'utilisation du mode « sans buffer » en lecture et en écriture et ne concernent que les accès utilisant le cache du système de fichiers (cf. section 3.2).

Ces deux modules implémentent donc tous les algorithmes que nous avons identifiés par ingénierie inversée [30] : lecture anticipée, écriture retardée, écriture immédiate, subdivision et concaténation des requêtes. Un aperçu de ces algorithmes est exposé plus loin.

### 4.1.5 L'ordonnanceur d'E/S
Ce module représente une file d'attente contenant les requêtes n'ayant pas encore été satisfaites par le disque. Ce module entre essentiellement en jeu quand l'application envoie des E/S asynchrones, c'est-à-dire qu'elle n'attend pas la satisfaction d'une requête avant d'en envoyer la suivante. Plusieurs politiques sont implémentées dans ce module : FCFS, SCAN, LOOK et plusieurs de leurs variantes.

**4.1.6 Le cache du disque**
Dans ce module sont décrits les différentes politiques de gestion de cache du disque que nous avons identifiées sur plusieurs machines différentes (interfaces SCSI et IDE). Ces comportements constituent une sorte de librairie permettant à l'utilisateur de décrire les politiques de cache de l'architecture à simuler : taille du cache, nombre et taille des segments, politiques de mise à jour, algorithmes de préchargement, d'écriture retardée et immédiate. Ces politiques ont été implémentées d'après les tests effectués sur plusieurs machines et d'après des politiques décrites dans la littérature [10] [17].

**4.1.7 Le disque**
Ce module simule toutes les latences mécaniques causées par les accès en lecture/écriture sur le disque. Nous prenons en compte les paramètres suivants : le nombre de plateaux, le nombre de cylindres, le zonings (tailles des pistes pour chaque zones ainsi que la taille des zones [15]), le temps de changement de pistes, le temps de changement de têtes, « track skew », « cylinder skew », les secteurs de secours et leurs emplacements sur le disque, ainsi que les différents mappings possibles.

**4.2 Les stratégies du cache du système de fichiers**
Les politiques de cache du système de fichiers et celles du cache du disque sont décisives pour déterminer les performances d'E/S d'un système donné. Le comportement du cache du système de fichiers est spécifique à un mode d'accès donné et est le même pour les versions de système d'exploitation Windows et les systèmes de fichiers testées, excepté le mode « write through » qui peut différer d'après le système de fichiers utilisé en raison de la journalisation de NTFS.
Nous résumons dans ce qui suit le comportement du cache du système de fichiers décrit d'après les travaux que nous avons précédemment menés [30] [31].
Excepté le mode « sans buffer », à chaque mode d'accès défini dans la fonction *CreateFile* correspond un algorithme de préchargement (lecture ou écriture). Nous allons résumer ces comportements par rapport à un flux séquentiel en lecture et en écriture.

**4.2.1. Accès en lecture**
a) Mode « séquentiel » : pour ce mode tous les blocs de données demandés sont subdivisés ou concaténés par blocs de 64Ko. Les données sont préchargées séquentiellement par blocs de 64Ko. Le préchargement débute après la détection d'un ensemble de requêtes séquentielles (environ 3) et s'arrête à la fin du fichier ou quelques blocs (environ 2*taille de la requête) après l'arrêt de la réception des requêtes séquentielles. Pour des requêtes dont la taille n'est pas multiple de 64Ko, le comportement est différent et plus complexe. Les préchargements ne se font plus par un nombre fixe de blocs de 64Ko par requête. L'algorithme utilisé [32] ne sera pas explicité dans cet article mais il ressemble beaucoup à celui du mode « normal » (voir section suivante).
b) Mode « normal » : dans ce mode, pour un fichier lu séquentiellement, deux processus distincts accèdent aux données demandées par l'application : le *processus application* et *le processus système*. Lorsque l'on envoie la première requête (pour laquelle le préchargement commence) le processus système commence à charger les données demandées par l'application. Une fois la première requête satisfaite par ce processus (pour des tailles de requêtes supérieures à 64Ko), ce dernier saute une taille de requête et commence à précharger les données. Si l'on prend comme exemple des requêtes d'une taille de 256Ko, cette dernière est constituée de 4 blocs de 64Ko ($B_1$, $B_2$, $B_3$ et $B_4$), le système charge alors les quatre blocs, puis précharge le bloc $B_9$, $B_{10}$, $B_{11}$ et enfin $B_{12}$. Pendant ce temps l'application demande les blocs $B_5$, $B_6$, $B_7$ et $B_8$ qui seront traités par le processus application *en parallèle* avec ceux demandés par le processus système, ceci donne la séquence montrée dans la FIG.3.

**4.2.2. Accès en écriture**
a) Modes « séquentiel » et « normal » : les algorithmes utilisés par ces modes d'accès sont exactement les mêmes en écriture. Nous avons observé deux différents comportements qui dépendent de la taille des requêtes choisie : (i) le processus application n'écrit les données que sur le cache du système de fichiers et jamais sur le disque alors que le processus système vide le cache *progressivement* sur le disque, (ii) le processus application écrit les données sur le cache du système de fichiers *et* sur le disque et le processus système vide *périodiquement* ce cache. Dans le premier cas, le vidage du cache est effectué continuellement par un petit nombre de blocs de 64Ko en parallèle avec l'écriture sur le cache. Dans le second cas, le vidage est effectué périodiquement par un grand nombre de blocs d'une taille équivalente à la taille de l'espace de travail du processus moins une valeur constante :

o *Le processus application écrit uniquement sur le disque* : le processus système est le seul dans ce cas à vider les données sur le disque. Ce comportement se produit pour des tailles de requêtes de 128Ko, 256Ko ou des tailles inférieures ou égales à 96Ko. Le vidage est réalisé par blocs de 64Ko et effectué en parallèle avec la copie des données sur le cache du système de fichiers. Ce mode produit les meilleures performances en écriture pour les architectures testées [31]. Les raisons des chutes de performances pour les autres tailles de requêtes sont liées à l'allocation des *vues* (l'entité minimum de cache alloué à un fichier : 256Ko) du cache du système de fichiers et à l'algorithme de vidage de ce dernier qui ne sont pas optimisées pour des tailles de requêtes importantes.

o *Les processus système et application écrivent les données sur le disque* : pour un flux de requêtes d'écriture donnée, le processus application écrit périodiquement une partie des données de chaque requête sur le cache du système de fichiers et le reste sur le disque. C'est le processus système qui videra les données qui sont sur le cache vers le disque lorsque l'espace de travail du processus est rempli. L'algorithme selon lequel les blocs de données sont écrits par le processus application sur le cache ou sur le disque et celui selon lequel le vidage est effectué ont été identifiés [31] et implémentés dans notre simulateur. Si l'on considère, par exemple, des requêtes d'écriture d'une taille de 320Ko : pour la première requête, 3 blocs sont écrits sur le cache du système de fichiers et 3 sur le disque. Pour la deuxième requête 4 blocs sont écrits sur le cache et 2 sur le disque. Pour la troisième requête, 5 blocs sont écrits sur le cache et 1 sur le disque et pour la quatrième requête, les 6 blocs sont écrits sur le cache. Ceci nous donne une période de 4 requêtes de 320Ko. Si la taille de l'espace de travail est de 8Mo, le vidage des données sur le cache vers le disque effectué par le processus système se fait à chaque 7 à 8 requêtes de 320Ko. Pour plus de détails par rapport à cet algorithme, on peut consulter [32].

b) Mode « Write through » : lorsqu'une requête d'écriture est envoyée, les données sont d'abord copiées sur le cache du système de fichiers par blocs de 64Ko, puis sur le cache du disque et finalement sur le disque. Avant d'envoyer la requête suivante, le système modifie les meta-données qui correspondent au fichier écrit (table FAT sous FAT32 ou fichier log puis MFT sous NTFS) [32].
Tous les algorithmes succinctement décrits dans cette section ont été implémentés dans notre simulateur.

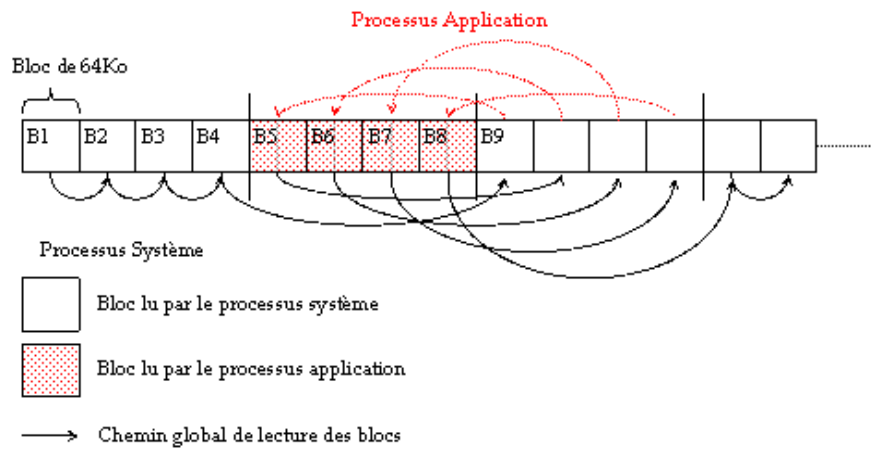

FIG. 3- Exemple de lecture d'un fichier en mode normal avec une taille de requête de 256Ko. La séquence de blocs demandés est vue du disque est : $B_1, B_2, B_3, B_4, B_9, B_5, B_{10}, B_6, B_{11}, B_7, B_{12}, B_8…$

## 4. 3 Configuration du simulateur

La configuration du simulateur peut être facilement effectuée sur un fichier texte. Trois ensembles de paramètres de configuration doivent être spécifiés :

a) L'architecture de stockage : on doit décrire dans cette partie les paramètres du sous système de stockage que l'on veut simuler : les caractéristiques mécaniques des disques, les algorithmes du cache du disque ou encore les mappings. Une petite partie de ces paramètres est fournie par le fabricant [8] [9], mais la plus grande partie peut être mesurée avec Wiotester, un outil développé dans ce but [28].
b) Le système d'exploitation : les paramètres de la configuration du système sont faciles à avoir, notamment en utilisant des outils libres tels que CacheSet [27].
c) L'application : deux cas sont ici possibles : l'utilisation des générateurs de requêtes et l'utilisation de traces réelles. Cet article ne concerne que le second cas. Si l'on choisit d'utiliser les traces réelles, ces dernières doivent être mesurées avec l'outil Filemon [23]. Ce traceur parait être l'outil le plus adéquat pour notre simulateur. Il montre néanmoins quelques défauts, comme le fait de ne pas dissocier les requêtes de l'application des requêtes du processus application ainsi que le fait de donner les noms de fichiers et le déplacement relatif au lieu de donner l'adresse logique de ce dernier. Ces défauts ont été corrigés par un programme que nous avons développé qui permet de détecter les différentes erreurs sur la trace produite par Filemon et d'en générer une nouvelle.

Les sorties du simulateur sont les performances du système en terme de taux de transfert moyen, temps de réponse unitaires (pour chaque requête générée). L'état de la simulation peut être visualisé à chaque instant pour chaque module grâce à l'interface graphique d'Omnet++ [6]. Nous pouvons ainsi consulter en temps réel des informations telles que le contenu de chaque mémoire, les vitesses des copies, les temps de traitement au niveau de chaque module, l'état du disque, les requêtes traitées ainsi qu'une multitude d'autres paramètres.

**4.4 La validation du simulateur par traces réelles**
**4.4.1 Méthodologie de validation du simulateur**
Nous nous intéressons dans cette section à la validation du simulateur par rapport à des traces réelles d'E/S et non plus à des charges de travail modélisées [29] qui se rapprocheraient des traces réelles.

Les applications à E/S intensives évoluent de plus en plus vite et il devient de plus en plus difficile de concevoir et d'implémenter des outils de benchmarking réalistes et qui sont représentatifs de ces applications. En revanche, les outils de benchmarking synthétiques sont un bon compromis, ils ne permettent pas de mesurer les performances d'une application réelle mais donnent une certaine flexibilité quant à la définition des charges de travail à imposer.

Nous avons utilisé deux outils de benchmarking pour la validation de notre simulateur : l'outil Sqlio [25] développé par Microsoft et Wiotester développé par nous-mêmes [28] dans le but de corriger certains des défauts des outils de benchmarking synthétiques actuels (Sqlio, Iometer[26], iozone [24]).

L'outil Sqlio permet de tester plusieurs charges d'E/S différentes en générant des requêtes suivant une configuration déterminée par l'utilisateur. Cette configuration est très limitée par rapport à ce qu'offre notre simulateur, mais reste riche par rapport aux autres outils de benchmarking. Sqlio présente néanmoins plusieurs défauts tels que l'impossibilité de tester la totalité des modes d'accès proposés par le système Windows et le fait de créer des fichiers de test pouvant être extrêmement fragmentés ce qui altère fortement les performances.

Le deuxième outil de benchmarking utilisé est Wiotester. Il offre moins de possibilité quant à la configuration de la charge d'E/S appliquée mais corrige tous les défauts de Sqlio par rapport aux modes d'accès utilisés et à la fragmentation des fichiers.

La procédure de validation de notre simulateur consiste à exécuter dans un premier temps l'outil de benchmarking Sqlio ou Wiotester sur une architecture donnée pendant environ une minute en mesurant les performances puis à récupérer la trace générée par ces outils. Nous injectons par la suite la trace récupérée dans notre simulateur après avoir configuré les paramètres liés à l'architecture de stockage et au système d'exploitation. Après avoir lancé la simulation, nous récupérons les performances simulées et les comparons avec celles mesurées pour toutes les architectures et les configurations de charges d'E/S testées. La métrique de performance prise en compte pour notre étude est la somme des temps de réponse de la totalité des requêtes envoyées pour une expérience donnée. Nous nous sommes assurés avant la réalisation des tests que le fait de tracer les E/S n'ait qu'un impact réduit sur les performances d'E/S.

Les différents tests exécutés dans le cadre de cette validation sont les suivants :
   1. Pour les opérations de lecture :

o   Accès séquentiel en mode « sans buffer »
            o   Accès séquentiel en mode « normal »
            o   Accès aléatoire en mode « sans buffer »
            o   Accès aléatoire en mode « normal »
    2.  Pour les opérations d'écriture :
            o   Ecriture en mode « normal » (même comportement que le mode séquentiel)
            o   Ecriture en mode « sans buffer »

Pour les opérations de lecture, les tests ont été réalisés sur des fichiers fragmentés et défragmentés. Les résultats de ces deux types de test étaient équivalents. Pour l'écriture, ces deux types de tests ne sont pas réalisables, l'écriture se faisant d'après l'espace disponible sur le disque. Les tailles de requêtes testées varient entre 32Ko et 512Ko. Davantage de tailles de requêtes ont été testées pour les écritures afin d'illustrer les différents phénomènes que nous avons exposés plus haut (cf. section 4.2.2).

Nous avons choisi de montrer les résultats de validation sur les trois architectures exposées sur TAB.1. Chaque mesure et simulation ont été effectuées au minimum trois fois et la moyenne des valeurs a été prise en compte.

|  | Config 1(Dell) | Config 2(HP) | Config 3(Asus) |
|---|---|---|---|
| Processeur | Intel Pentium4 1,8GHz | Intel Pentium 3 500MHz | Intel Centrino 1.6 GHz |
| Memoire (Mo) | 256 | 192 | 512 |
| Disque | Fujitsu MAN 3184MP | Toshiba MK6012MAP | Hitachi Travelstar 80GN |
| Capacité (Go) | 18.4 | 6 | 60 |
| Vitesse de rotation (tpm) | 10000 | 4200 | 4200 |
| Avg-seek (ms) r/w | 4.5 / 5 | 13 | 13 |
| Min-max seeks (r/w) | 0.4/0.6 – 11/12 | 3-24 | 2.5-31 |
| Taille des pistes (Ko) | 221-377 | - | 224-434 |
| Taille du cache disque (Mo) | 8 | 1 | 8 |
| Interface | Ultra 160 SCSI | ATA 4 | ATA 6 |
| Version Windows | 2000 Professionel SP4 | 2000 Professionel SP4 | XP edition Familliale SP2 |
| Système de fichiers | FAT32/NTFS | NTFS | FAT32/NTFS |

**TAB. 1** – Caractéristiques des configurations testées

**4.4.2 Résultats de la validation**

Les résultats de la simulation sont représentés dans la FIG. 4. Ils sont exprimés en terme de pourcentage d'erreur entre les résultats des benchmarks et des simulations des traces d'E/S.
Le pourcentage d'erreur est calculé comme suit :
%Erreur = 100*abs(temps mesuré – temps simulé)/(temps mesuré)

**Les opérations de lecture**

Nous pouvons remarquer pour la config2 et la config3 (cf. TAB.1) pour les accès aléatoires et séquentiels en mode « sans buffer » et « normal » que le pourcentage d'erreur observé pour la simulation par rapport à la mesure est de moins de 9% dans tous les cas et que la moyenne de cette erreur est d'environ 6%. Ceci démontre parfaitement la validité du simulateur pour ces deux architectures. En revanche, pour la config1, nous remarquons un pourcentage d'erreur avoisinant les 45% pour la lecture séquentielle en mode « normal » et en mode « sans buffer » pour des requêtes de 128ko. Après avoir effectué plusieurs expériences et simulations, nous avons identifié la cause de cette variation. En effet, le disque avait un comportement irrégulier pour les requêtes de 128ko en mode « sans buffer ». Ceci était dû au fait qu'il préchargeait des blocs de 512Ko dans le cache et qu'après les avoir envoyés vers la mémoire (par tranche de 128ko), il était obligé de se repositionner sur le bon bloc avant de poursuivre la lecture (en perdant environ un tour de piste), alors que pour les autres tailles de requêtes, le disque chargeait les données dans le cache et les envoyait vers la mémoire au fur et à mesure. Nous avons évalué cette perte et avons constaté qu'elle engendrait une chute de performance de 50% par rapport à ce que devait produire le disque, d'où l'erreur au niveau du simulateur dans la FIG. 4-1. Pour la FIG. 4-2, nous remarquons des différences de performances variant entre 25% et 42%., ceci est aussi dû à un préchargement effectué par le cache du disque. Ce préchargement de 512Ko se produit à chaque fois que deux blocs non éloignés (blocs locaux) sur le disque sont demandés. Dans la FIG. 3 par exemple, nous remarquons la suite de blocs : « …$B_4$, $B_9$, $B_5$… », dès que le contrôleur détecte la suite B$_4$, B$_9$, B$_5$, il précharge à partir de B$_5$ un bloc de 512Ko [31]. Une fois ces comportements implémentés, l'erreur entre les simulations et les mesures s'est réduite à des valeurs toujours inférieures à 10% (cf. FIG. 4 -2). Ce type de comportement n'est valable que pour ce type de disque et il a été implémenté pour n'être pris en compte que pour ce type de disque.

**Les opérations d'écriture**

Concernant les opérations d'écriture, nous remarquons que les taux d'erreurs ne dépassent jamais les 10% avec une moyenne d'environ 6% et ceci malgré le fait que l'état initial du contenu de la mémoire, celui du cache du disque et l'état initial du disque n'aient pas été pris en compte. Les fichiers écrits étaient fragmentés, ce qui rend les simulations effectuées encore plus pertinentes. En effet, ceci a permit une simulation plus réaliste des E/S.

**4.5 Quelques améliorations à apporter**

La vitesse du simulateur par rapport aux mesures est relativement lente, la simulation est en moyenne trois à quatre fois plus lente que la mesure. La simulation des modules mémoires est la plus lente, en particulier les mises à jour : le cache du disque par exemple est mis à jour secteur par secteur. Nous avons utilisé les classes de file d'Omnet++ pour l'implémentation de ce type de module. La solution ce problème consiste à développer de nouvelles librairies plus rapides que celles d'Omnet++ permettant donc de la remplacer. Nous pensons ainsi facilement arriver à rendre la simulation deux fois moins rapide (voire mieux) que la mesure dans un premier temps.

Un autre problème, la propagation d'erreur au niveau des temps de réponse, a été rencontré lors de nos simulations. En effet, si pour une requête, par exemple, le temps de réponse produit par notre simulateur $T_s$ est inférieur à celui mesuré $T_m$ et que la requête suivante est séquentielle et si le temps d'inter arrivée entre les requêtes est nul, la nouvelle requête ne sera envoyée qu'après ($T_m$- $T_s$) sur le simulateur (la rotation du disque se poursuit pendant ce temps). Ceci cause une latence de rotation du disque supplémentaire par rapport au simulateur qui été inexistante lors de la mesure. Si le temps d'une rotation complète est de $T_r$, cette latence sera $T_r$ - ($T_m$- $T_s$). Nous avons résolu ce problème en intégrant une tolérance par rapport aux temps de réponse permettant de ne pas répercuter les erreurs.

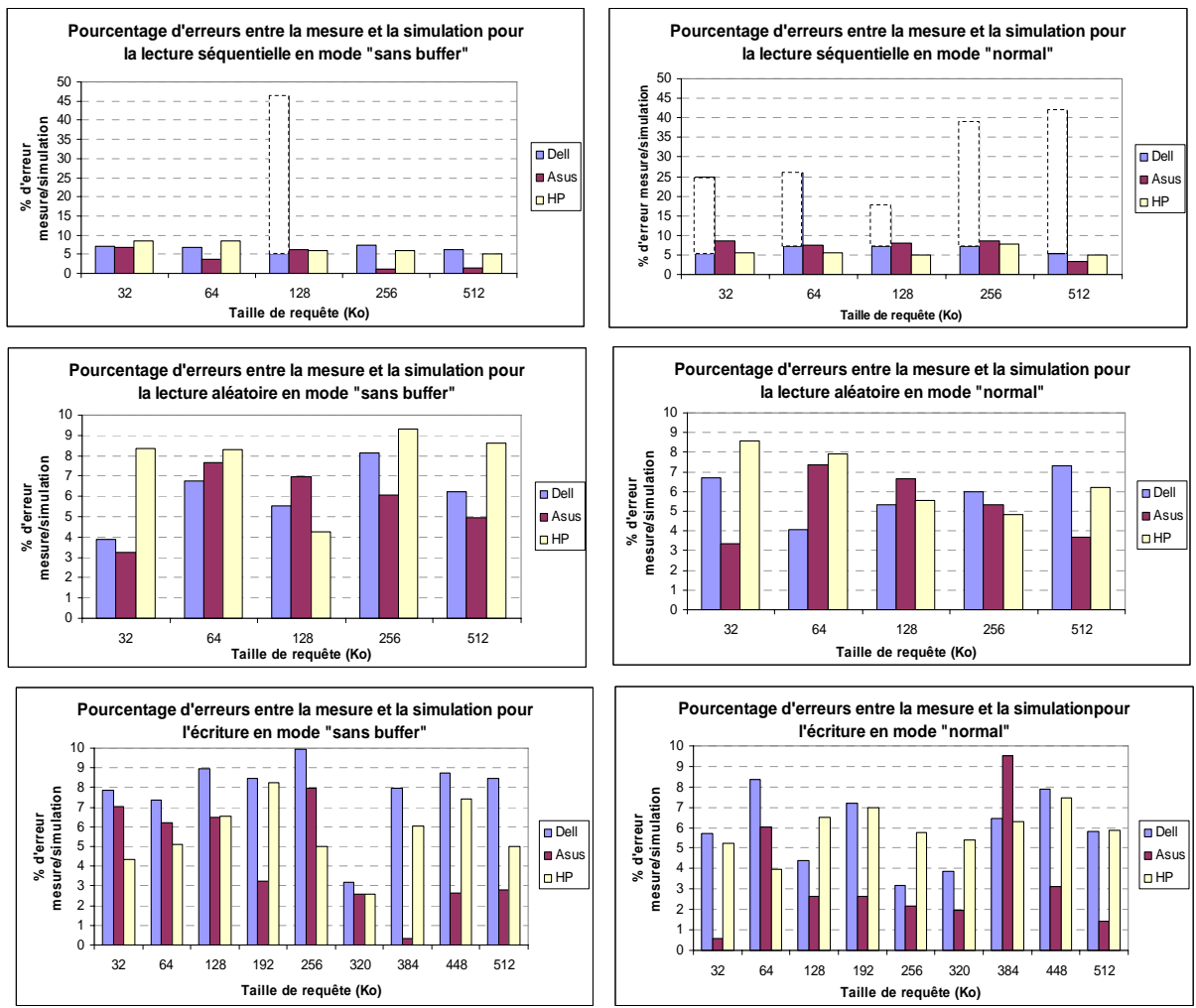

**FIG. 4**- Résultats de la validation du simulateur d'E/S avec Sqlio et Wiotester : % erreur entre les simulation et la mesure pour 1) la lecture séquentielle en mode « sans buffer », 2) la lecture séquentielle en mode « normal », 3) la lecture aléatoire en mode « sans buffer », 4) la lecture aléatoire en mode « normal », 5) l'écriture en mode « sans buffer » et 6) l'écriture en mode « normal ».

**5. Résumé et perspectives**

Nous proposons dans cet article un simulateur d'E/S sur PC très précis et très flexible permettant de simuler des traces réelles d'E/S avec une erreur moyenne de 6% sur le temps de réponse global.
Ce simulateur permet à un utilisateur d'identifier la meilleure stratégie d'accès aux fichiers pour une application d'après ces contraintes d'E/S. Il permet aussi de réaliser très rapidement un meilleur choix à moindre coût quant à l'architecture à utiliser par rapport à une charge d'E/S donnée.
Ce simulateur permet de résoudre le problème de la représentativité des outils de benchmarking par rapport aux applications. En effet, les outils de benchmarking ne permettent qu'une flexibilité très limitée quant à la charge d'E/S à mesurer, en plus du fait qu'ils ne prennent pas en compte toutes les subtilités de l'utilisation des modes d'accès fournis par le système d'exploitation.
Le simulateur décrit dans cet article à été construit sous Omnet++, un environnement de simulation libre, très flexible et pourvu d'une très bonne interface graphique permettant de concevoir des simulateurs basés sur des techniques de passage de messages.
Tous les éléments entrant en jeu lors de la satisfaction d'une requête d'E/S sont pris en compte dans le simulateur :
- La configuration détaillée des traces d'E/S,
- les paramètres liés au système d'exploitation (cache du système de fichiers et mémoire processus),
- les paramètres de configuration de l'architecture de stockage (interface, cache du disque et disque).

La configuration d'un simulateur n'étant pas toujours facile à réaliser, un outil d'extraction de paramètres de stockage [28] très simple à utiliser est disponible.
Les résultats de la simulation sont, entre autres, les temps de réponse unitaires et les taux de transfert. L'évolution des différents paramètres peut être visualisée pendant la simulation.

Le test d'autres architectures est en cours dans le but d'enrichir la librairie des différents comportements des disques. Nous projetons d'optimiser les files sous Omnet++, ceci permettra d'effectuer des simulations beaucoup plus rapides.
Nous avons entamé l'analyse des performances d'E/S des accès aux fichiers stockés sur des machines distantes. L'intégration des résultats dans le simulateur pour permettre des simulations incluant des machines distantes nous parait extrêmement intéressante. A moyen terme, la construction d'un simulateur incluant plusieurs architectures différentes sur un réseau dédié permettra d'entamer les réflexions sur les stratégies de stockage sur des grilles hétérogènes.

**6. Disponibilité du logiciel**

Le simulateur et son code source seront librement disponibles sur : http://www.prism.uvsq.fr/~jboukh.

**Bibliographie**